\documentclass[prl,twocolumn,floatfix,aps]{revtex4}
\usepackage[pdftex]{graphicx}
\usepackage[utf8]{inputenc}
\usepackage{bm}
\usepackage{color}
\usepackage{amssymb}
\usepackage{mathrsfs}
\usepackage{amsmath}

\newcommand{\newc}{\newcommand}

\newc{\be}{\begin{equation}}
\newc{\ee}{\end{equation}}
\newc{\bea}{\begin{eqnarray}}
\newc{\eea}{\end{eqnarray}}
\newc{\ie}{{\it i.e.} }
\newc{\eg}{{\it e.g.} }
\newc{\etc}{{\it etc.} }
\newc{\etal}{{\it et al.} } 
\newc{\ra}{\rightarrow}
\newc{\lra}{\leftrightarrow}
\newc{\lsim}{\buildrel{<}\over{\sim}}
\newc{\gsim}{\buildrel{>}\over{\sim}}

\begin{document}

\title{Comment on Phys.Rev. Lett. {\bf 122}, 084501 (2019) by A. Esposito, R. Krichevsky and A. Nicolis}
\author{C. Tannous}
\affiliation{LABSTICC - UBO, UMR-6285 CNRS, Brest, F-29200 FRANCE 
\thanks{Tel.: (33) 2.98.01.62.28,  E-mail: tannous@univ-brest.fr}}
\author{J. Gieraltowski}
\affiliation{Laboratoire Géosciences Océan - IUEM, UMR-6538 CNRS, Plouzan\'e, 29280 FRANCE}

\maketitle

In Phys.Rev. Lett. {\bf 122}, 084501 (2019), Esposito \etal introduce the notion of a
gravitational mass $M$ carried by sound waves as given by:

\be
M=-\frac{d \log c_s}{d \log \rho_m}\frac{E}{c_L^2}
\ee

where $c_s$ is sound speed for the (longitudinal $L$ or transverse $T$) type of sound wave considered.  
$E$ is sound wave energy and $\rho_m$ the mass density of the wave carrier medium.

They apply this notion to earthquakes and deduce that for a Richter scale seism of magnitude $R=9$
there is an induced change $\delta g=10^{-4}$ nm/s$^2$ in gravitational acceleration,  
concluding that presently $\delta g$ is too small to be measured. However, they point out that 
it might be possible soon to detect these changes since a considerable progress has been achieved recently with the detection of gravitational waves with the LIGO experiment~\cite{LIGO}.

Surprisingly, Esposito \etal have, in fact, hinted at an acoustic version of Einstein formula
$E=M c_s^2$ with $E$ the quake energy when one approximates the $-\frac{d \log c_s}{d \log \rho_m}$ 
factor by 1.

The energy $E$ produces a mass variation $\delta M_\oplus$  
inducing a local change  $\delta g=\frac{G \delta M_\oplus}{R_\oplus^2}$ in the 
gravitational acceleration $g$, with $R_\oplus, M_\oplus$ Earth radius and mass respectively. 

An earthquake releases energies through various channels such as heat, radiation or
frictional weakening and inelastic deformation processes in rocks leading e.g. to formation of cracks.
Presently, the characterization of seisms is based on moment-magnitude scale~\cite{Bormann}
which is more accurate than the Richter scale especially in the case of strong earthquakes.

\begin{figure}[htbp]
\begin{center}
\includegraphics[angle=0,width=9.5cm,clip=]{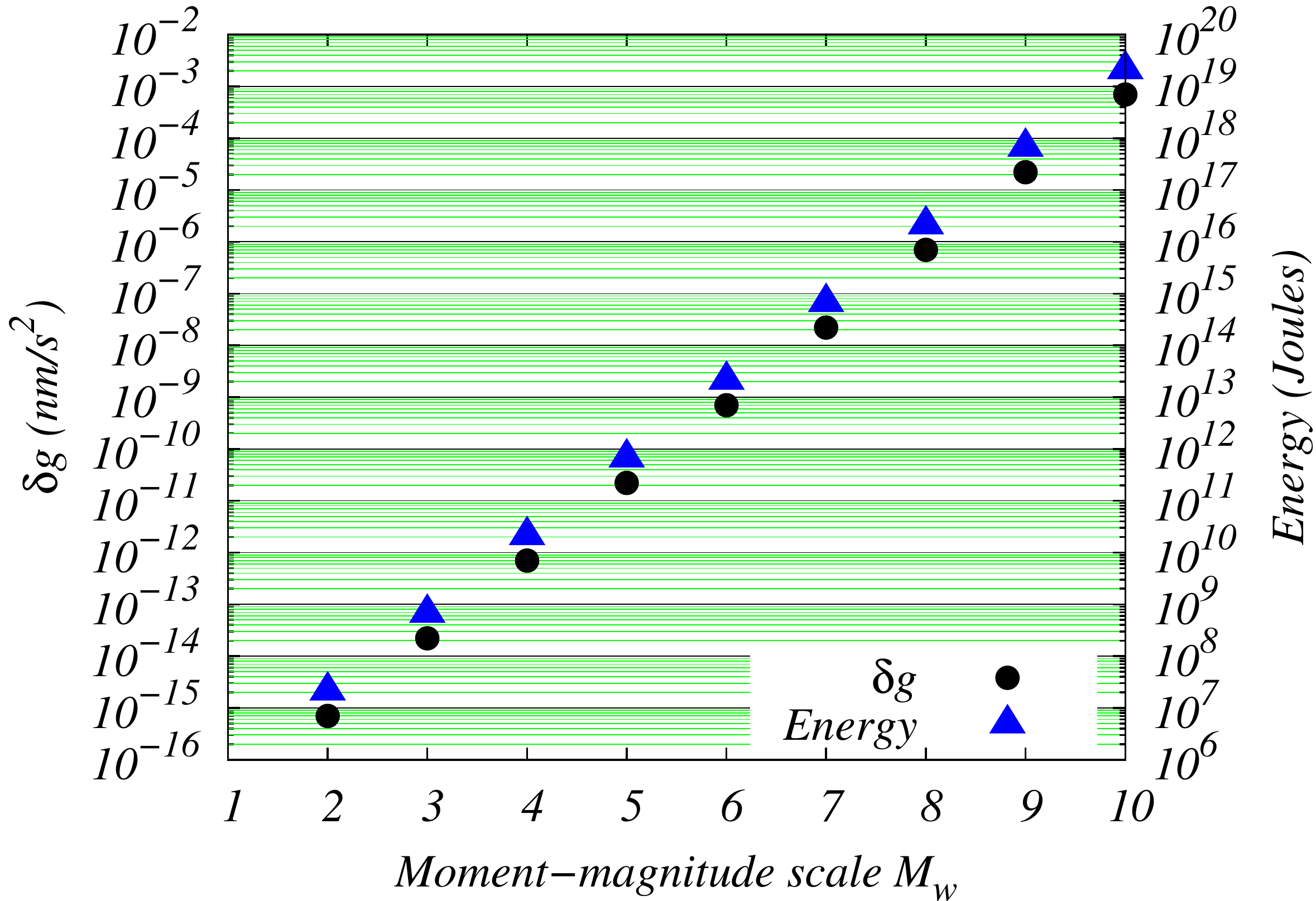}
\end{center}
 \caption{To the left, variation of the gravitational acceleration induced by a seism of magnitude
$M_w$ on the moment-magnitude scale and to the right variation of seism energy versus $M_w$. }
\label{moment}
\end{figure}

In the Richter scale, quake energy is obtained from the Gutenberg-Richter~\cite{Bormann} 
formula $E~=~10^{1.5 \times R + 4.8}$, while in the moment-magnitude $M_w$ scale, it is given by
$E~=~10^{1.5 \times M_w + 4.32}$. This allows us to plot simultaneously versus  $M_w$, 
the gravitational acceleration change and seism energy as displayed in fig~\ref{moment}.

According to the graph, the strongest quake ($M_w$=10) produces 
$\delta g=10^{-3}$ nm/s$^2$ slightly below the current sensitivity $10^{-2}$ nm/s$^2$ of 
superconducting gravimeters. 

By comparison, the Sumatra-Andaman quake~\cite{Sumatra} is one of the strongest ever recorded over the last 65 years. It actually occurred in the Indian Ocean on December 26, 2004,  
releasing approximately 1.1 $\times 10^{17}$ joules as obtained from moment-magnitude 
scale~\cite{Bormann} analysis.  
It corresponds on our graph to  $\delta g=10^{-5}$ nm/s$^2$. 
While this value is beyond current sensitivity, there are several possible 
avenues to attain it soon  such as Strontium optical lattice clocks,
Cold atom or laser interferometric techniques inspired from LIGO experiment. \\

{\bf Acknowledgments}: Discussions with Prof. Jacques Déverchère were very helpful to 
put this work into perspective.

\end{document}